\documentclass[twocolumn,showpacs,preprintnumbers,amsmath,amssymb, nofootinbib]{revtex4}
\usepackage{pdflscape}
\usepackage{color}
\usepackage{dcolumn}% Align table columns on decimal point
\usepackage{bm}% bold math
\usepackage{epsfig}%
\usepackage{amsmath}
\usepackage[colorlinks]{hyperref}
\usepackage{ulem}
\begin{document}
%\lipsum

\title{The Second Law, Symmetry of Time Reversal and Thermodynamic Equilibrium in Superconductors.}% Force line breaks with \\

\author{Vladimir Kozhevnikov}
\affiliation{Tulsa Community College, Tulsa, Oklahoma 74119, USA}.\\


\begin{abstract}
\noindent 
The Second law of thermodynamics %, universally applicable to all states of all sorts of matter and radiation, 
occupies a special place among the laws of nature. % , is undeniably the brightest jewel in the tiara of   laws of thermodynamics.
Suffice to recall that this law establishes the arrow of time and controls all processes associated with entropy; it  underlies %the principle of least action, 
atomistic structure of matter, quantization of energy, and much more. However, there is a significant gap concerning the  Second law as it applies to magnetizing materials, especially to superconductors. This communication is targeted to fill the gap. The concepts of equilibrium and reversibility in thermodynamic processes are reviewed with particular attention to specifics % features 
of superconductivity. The most important of these is the zero entropy of samples in a state of thermodynamic equilibrium, that is, the ideal ordering of microscopic currents formed by conduction electrons bound into Cooper pairs. The uniqueness of such an ordering, justified by thermodynamics and confirmed by experiment, is that superconductors are the only %(and readily accessible!) 
conducting materials satisfying the Clausius-Planck theoretical limit of absolute reversibility of thermodynamic processes, which means that properties of superconductors are not affected by time and, consequently, are not subject to aging. %It is shown that 
%In turn, zero entropy leads to all the ``miracles" of the superconducting state, such as perfect diamagnetism, zero resistance to a subcritical total current, etc., which with necessity steam from the Second law.  Conditions for achieving the state  of equilibrium in superconductors are framed out.  Awareness of these aspects is important for advancing researches in the field of superconductivity and related phenomena.
In turn, zero entropy explains %leads to 
%all the “miracles” 
the “miraculous” properties of the superconducting state, in particular zero resistance to a subcritical total current, which, as it is shown, is a consequence of the Second law. Conditions for achieving the state  of equilibrium in superconductors are framed out. The aspects discussed are important for advancing researches in the field of superconductivity and related phenomena, and will also be of interest to readers curious in thermodynamics.

%will be of interest to readers who are keen on thermodynamics and their awareness} is important for advancing researches on superconductivity and related phenomena. 

%Awareness of these aspects \tr{will be of interest to readers who are keen on thermodynamics and} is important for advancing researches on superconductivity and related phenomena.

%These include (1) the primary role of the field strength $\textbf{H}$ in forming magnetization; (2) the zero entropy of samples in a state of thermodynamic equilibrium; and (3) two kinds of superconducting currents in samples out of equilibrium. The most important of these three is zero entropy, which implies an ideal ordering of the field-induced microscopic currents formed by conduction electrons bound in Cooper pairs. 

\end{abstract}

%\pacs{74.20.-z, 74.25.Ha, 78.70.Nx}% PACS, the Physics and Astro\Deltanomy
\maketitle  

Introduction. 
\vspace{2mm}

By the mid 1800s it became clear that the First fundamental principle of thermodynamics, or the First law, representing the energy conservation, was insufficient to explain the absence of time reversal symmetry (TRS) in thermal processes, that is irreversibility of macroscopic properties of statistical systems whose microscopic properties are reversible.  %\cite{Thompson(Kelvin)}. 
The solution was found by Clausius, who introduced the Second %fundamental principle, or the Second 
law of thermodynamics \cite{Clausius-1867}, based on a previously unknown physical quantity that he named entropy. The nature of entropy as a measure of probability was disclosed by Boltzmann \cite{Boltzmann}, who thereby discovered a qualitatively new aspect of classical thermodynamics: the discreteness of entropy and, consequently, of all related properties. An enormous fruitfulness  of the Second law in Boltzmann's interpretation for virtually all domains of physical  science was demonstrated by Planck  \cite{Planck_8 lecturs}. Prominently, this led him to solve the problem of radiant heat and discover two physical constants, one of which Planck named after Boltzmann, and the other known as Planck's constant. This was a major triumph for classical thermodynamics.    

%In particular, it helped him solve the problem of radiant heat and discover two physical constants, one of which Planck named after Boltzmann, and the other known as Planck's constant. Needless to recall that a profound understanding of entropy led Planck to discover two most fundamental physical constants, one of which was named by Planck after Boltzmann, and the other known as the Planck constant.
%Recall that a deep understanding of entropy led Planck to discover two most fundamental physical constants, one of which was named after Boltzmann, and the other known as the Planck constant; this  was a major triumph for classical thermodynamics.      

The Second law can be verbalized in various forms, see, e.g., \cite{Clausius-1867,Thompson(Kelvin),Planck,Jarzynski}. %Planck's formulation reads \cite{Planck} 
The shortest  one, due to Planck \cite{Planck}, reeds
``The total change of the entropy\,$\geqq$\,0", where the inequality relates to irreversible processes and the equality refers to reversible ones. At the same time, an exact zero or completely reversible processes were considered as a theoretical limit, unattainable in reality \cite{Clausius-1867,Planck,Callen}.

The Second law, originally formulated as a generalization of empirical facts, was rigorously substantiated by statistical thermodynamics \cite{Gibbs,Schrodinger}. The validity of the law has been proven for all sorts of matter in any aggregate state, homogeneous and inhomogeneous systems and radiation \cite{Planck-radiation,Planck_8 lecturs,LL-StatPhys,Callen}; the latest addition to this list is the information processing  \cite{Landauer,Dagoa}. Overall, the Second law is a fundamental law of nature, which ``will never be overthrown" (Einstein about thermodynamics\footnote{Complete quote: ``A theory is the more impressive the greater the simplicity of its premises is, the more different kinds of things it relates, and the more extended is its area of applicability. Therefore the deep impression which classical thermodynamics made upon me. It is the only physical theory of universal content concerning which I am convinced that, within the framework of the applicability of its basic concepts, it will never be overthrown (for the special attention of those who are skeptical on principle)"\cite{Einstein}.}).    

Yet, there is a significant gap concerning the Second law as it applies  to magnetizing materials\footnote{Magnetizing materials are such whose magnetization vanishes in the absence of an external magnetic field.}, especially to superconductors. The objective of this communication is to fill this gap. 

To begin, let us look closer at the quoted Planck formulation of the  law and revisit the concept of thermodynamic equilibrium in reversible processes, that is, in successions of  equilibrium states with constant entropy. 

By definition, the equilibrium state of a statistical system is a state, whose properties are determined by the state parameters (e.g., volume, temperature, applied field, and the like), %\textit{etc.}), 
regardless on the path by which the system is brought to that state. Consequently,  the properties of systems in equilibrium are reproducible and time independent,  while entropy takes the maximum of the accessible values in both reversible and irreversible processes.
The difference between the two is that 
 in the former, the permutation of the final and initial states, i.e., time reversal (TR), \textit{completely} restores the initial state of the system{\footnote{A complete restoration implies restoration without any traces of the past, which means that the system is not subject to aging.}, whereas in the latter this is not the case, due to the difference in the entropy of the initial and final states in irreversible processes.  
Thus, systems in reversible processes possess TRS, whereas in irreversible processes this symmetry is absent. In other words, the concepts of \textit{reversibility and TRS are equivalent}: both can, in principle,  materialize only together and only in strictly adiabatic processes. The same follows from the fact that the time differential enters in the equations of reversible processes only as an even power, which includes the zero power for properties unrelated to time \cite{Planck_8 lecturs}.  

On the other hand, properties of statistical systems that are out of equilibrium 
are mostly determined by transport processes caused by the exchange of mass, charge, energy, linear and angular momentums. Such systems have a “memory” of their recent history, meaning that in cyclical processes they exhibit hysteresis. For nonequilibrium systems, the Second law asserts that  they   \textit{monotonically evolve} toward equilibrium and therefore do not possess TRS by definition.   

The above is of a universal nature and is well 
covered in textbooks on thermodynamics. %, e.g., \cite{Callen}. 
However, to our knowledge, a discussion of superconductivity from the viewpoint of the Second law is  missing. Meanwhile, as can be seen from the abstract, 
properties of superconductors are clearly fall under the scope of the Second law and therefore its application can be beneficial for understanding superconductivity and related phenomena.
%  the application of this law  can be beneficial for understanding the nature of superconductivity. %this phenomenon. % of superconductivity. %no less fruitful than it was in relation to other phenomena discussed by Planck. 
An endeavor of such a discussion is presented here. %below.
%Below is an endeavor of such a discussion.
 %\cite{Planck_8 lecturs}. 

The paper consists of five sections. The first section, following the introduction, is devoted to the origin and mechanism of the time-symmetrical magnetization in normal diamagnets and superconductors. The zero entropy of these materials and its implications for equilibrium properties of superconductors are discussed it the second section. The next section concerns non-equilibrium properties of superconductors caused by total currents; in particular, it is shown that the zero resistivity of superconductors stems from the Second law. 
Conditions for achieving thermodynamic equilibrium in superconductors are addressed in the fourth section. 
The final section contains a summary.   %A summary is provided in the fifth section.    

%in many publications of recent years, the authors claim to have observed \textit{breaking} of TRS in superconductors. Since the entropy of superconductors is constant (zero), such claims contradict the Second law and, therefore, are unrealistic. At the same time, there are specific features concerning the Second law  as applied to superconductors, which are insufficiently articulated in literature. Non-acquaintance with these features is a possible reason of statements similar to TRS breaking. These features are discussed below.% Unless otherwise is indicated,  the following is related to equilibrium states.
\vspace{2mm} 

1. Time reversal symmetry in diamagnets. 
\vspace{1mm}

TRS rises no questions regarding electrically polarizing materials \cite{LL-electrodynamics}, since polarization is unrelated to time. However, this is not the case for diamagnets, because magnetization in these materials is caused by induced microscopic currents persistently circulating within the sample.
%Magnetization of diamagnets, including superconductors, is caused by microscopic currents persistently circulating within the sample. 
Therefore, TR (the  transformation $t \rightarrow -t$) reverses direction of these currents\footnote{In the current-carrying systems, time reversal ($t\rightarrow -t$) is equivalent to reversal of the sign of charge carriers  ($e \rightarrow -e$).} and, accordingly, the direction of the magnetic moments caused by them. However, if this is the case, than TR should reverse magnetization at the unchanged state parameters, which conflicts with the Second law. %, since the entropy was not altered.

To sort out with this issue, we need to refresh the notions of induction $\textbf{B}$ and intensity (or strength)   $\textbf{H}$ of the magnetic field, connected via  $\textbf{B}=\textbf{H}+4\pi \textbf{I}$, where $\textbf{I}$ is magnetization, magnetic moment per unite volume induced by the field. 
After Lorentz \cite{Lorentz}, $\textbf{B}$ is a mean field due to all microscopic currents in the sample. In turn, $\textbf{I}$ is related to the field through the \textit{fundamental equation of induced magnetism} \cite{Maxwell}, which reads
\begin{equation}
	\textbf{I}=\chi\textbf{H},
 \end{equation}   
where $\chi$ is the volume magnetic susceptibility or coefficient of induced magnetization. 

Therefore, the intensity $\textbf{H}$, also called the magnetizing force \cite{Maxwell}, controls magnetization: it causes magnetic moments in the microscopic units composing the sample. 
In other words,  $\textbf{H}$ is a mean field acting on a microscopic unit; it does not include the field due to the unit in question, while $\textbf{B}$, the mean field due to all units, includes it. This seemingly minor difference between $\textbf{H}$ and $\textbf{B}$ fields inside the sample (outside $\textbf{B}\equiv \textbf{H}$) leads to principally different boundary conditions for these fields \cite{LL-electrodynamics}. Accordingly, $\textbf{H}$ and $\textbf{B}$ \textit{must be carefully distinguished} \cite{Maxwell}. Besides, in superconductors the difference is by no means minor: in the \texttt{S} (superconducting) phase $\textbf{B}=0$, while $\textbf{H}$ exceeds or equals the applied field $\textbf{H}_0$. For example, in spherical superconductors in the Meissner state $\textbf{H}=1.5\, \textbf{H}_0$. \textbf{B} can be directly measured using an external magnetic probe(s) implanted into the sample, such as muons in $\mu$SR spectroscopy. In turn, \textbf{H} can be measured using intrinsic magnetic moments, as is done, e.g., in ESR and NMR spectroscopy\footnote{In the intermediate state of superconductors, $\textbf{H}$ can be measured by $\mu$SR with the use of boundary conditions \cite{Egorov,VK_book}.}.

The $\textbf{H}$ field, as any other magnetic field, is solenoidal\footnote{This fact does not exclude a formal possibility of using the scalar potential for $\textbf{H}$. See \cite{VK_book} for details.}. % \cite{Tamm,Maxwell,VK_book}.}. 
Therefore, its physically adequate description is based on a vector potential $\textbf{A}$ defined so as 
\begin{equation}
	\textbf{H}=\nabla \times \textbf{A}
\end{equation}
with zero divergence (i.e., $\nabla \cdot \textbf{A}=0$) presumed automatically\footnote{The condition $\nabla \cdot \textbf{A}=0$ follows from the absence of microscopic currents crossing the sample surface  
\cite{Tamm}. At the same time, $\nabla \cdot \textbf{A}=0$ implies that the linear momentum operator commutes with the vector potential taken in the circular gauge \cite{Landau_QM}.}   and the circular gauge for $\textbf{A}(=\textbf{H}\times \textbf{r}/2$, where $\textbf{r}$ is the radius-vector lying in the transverse to \textbf{H} plane), since this is the only gauge that complies with the cylindrical symmetry of
a uniform \textbf{H} field\footnote{The uniformity of $\textbf{H}$ is the necessary condition of thermodynamic equilibrium in magnetizing media.} (all directions in the planes transverse
to \textbf{H} are equivalent).
%\footnote{Regarding the vector potential of induction $\textbf{A}_B$, defined so as $\textbf{B}=\nabla\times \textbf{A}_B$, it is $\textbf{A}_B= \langle \textbf{A}\rangle=\mu_m\textbf{A}$, where $\langle \textbf{A}\rangle$ is the space average of $\textbf{A}$, and $\mu_m(=1+4\pi \chi)$ is magnetic permeability.} \cite{VK_book}. 
 
First, consider normal diamagnets. In these materials, at zero external field, the magnetic moments of bound electrons (i.e., moments caused by the orbital motion of electrons in atoms, ions, etc., as well as their spins) are compensated. The compensation is dictated by the Second law because the free  energy of an ensemble of non-magnetic units is less than that of the same units possessing magnetic moments. %\footnote{The principle of minimum free (Helmholtz and Gibbs) energy follows from the law of maximum entropy of equilibrium states.}. 
In turn, the spin compensation means that the bound electrons are effectively spinless. %\footnote{The same follows from the Pauli exclusion principle.}. 
There are also unbound conduction electrons, but their contribution into magnetic properties  
is paltry. 

When $\textbf{H}_0$ is switched on, the intensity inside the sample changes from zero\footnote{For simplicity, the terrestrial field is neglected. } to $\textbf{H}$ and each electron is subject to the Lorentz force $\textbf{F}_L$ in addition to the forces acting at zero field. At the absence of an applied electric field\footnote{In conductors, an external electric field %of not very high frequency ($\lesssim 10^9$ Hz) 
is screened by unbound electrons.} the Lorentz force is
\begin{equation}
	\textbf{F}_L=-\frac{e}{c}\left(\frac{\partial \textbf{A}}{\partial t}\right)+\frac{e}{c}\textbf{v}\times \textbf{H},
\end{equation}
where $\textbf{A}$ is the vector potential of $\textbf{H}$, $e$ is unit charge, $\textbf{v}$ is the electron velocity, $c$ is speed of light and $t$ is time.

The second term on the right hand side is the magnetic force $\textbf{F}_M$, which is directed perpendicular to $\textbf{v}$. Therefore, $\textbf{F}_M$ cannot change the kinetic energy of electrons and, accordingly, the state of the sample, since it does not do work. In other words, $\textbf{F}_M$   \textit{alone cannot} create magnetism \cite{Feynman_Lectures}.  

%The second term on the right hand side is the magnetic force $\textbf{F}_M$, which causes the electron's orbit (or the magnetic moment $\bm{\mu}$ due to the orbiting electron) to precess about the field $\textbf{H}$  with Larmor frequency $\bm{o}$ (omicron) $= -\gamma \textbf{H}$, where $\gamma(=e/2mc$ with $m$ equal to the free electron mass) is a gyromagnetic ratio of orbiting spinless electrons. This means that in a uniform   $\textbf{H}$-field all electron orbits precess synchronously, that is, without changing their mutual orientation, and parameters of the orbits (the orbital radius $R$ and the magnitude of magnetic moment $\bm{\mu}$ due to the orbiting) are unaffected by $\textbf{F}_M$\footnote{The parameters $R$ and $|\bm{\mu}|$ are unaltered providing the  frequency of the orbital revolution is much greater than the Larmor frequency. This condition is fulfilled by a giant margin in all magnetics \cite{VK_book}.}. Hence, $\textbf{F}_M$ does not change the state of the sample in consistency with the fact that it does not do work. In other words, $\textbf{F}_M$   \textit{alone cannot} create magnetism. 

The magnetization of diamagnets is due to the first term in Eq.\,(3). This is the electric force $\textbf{F}_E$ caused by the induced vortex electric field existing while the magnetic field is changing. As one can show \cite{Tamm,VK_book}, in a result of its action the angular velocity of each orbiting electron changes by $\bm{\omega}_i=-(e/2mc)\textbf{H}$, i.e., the change $\bm{\omega}_i$ is equal to the Larmor frequency $\bm{o}$\,(omicron) $=-(eg/2mc)\textbf{H}$, where $g=1$, as it should be for the spinless orbiting electrons, and $m$ is the free electron mass. %where $m$ is the free electron mass and the $g$-factor of the spinless orbiting electrons is unity. 

Thus, every bound electron acquires an induced magnetic moment $\bm{\mu}_i$ equal to
\begin{equation}
	\bm{\mu}_i=\frac{\pi r_i^2}{c}\,\frac{e}{2\pi}\bm{\omega}_i=\frac{\pi r_i^2}{c}\,\frac{e}{2\pi}\bm{o}=-\frac{e^2r_i^2}{4mc^2}\textbf{H},
\end{equation}
where $r_i$ is the rms radius of the induced currents\footnote{Note the identity of Eq.\,(4) with the classical Langevin formula derived without the first term in Eq.\,(3). This derivation is considered incorrect since it does not explain the energy source of the induced magnetism \cite{Feynman_Lectures}.}. 

After that the bound electrons continue precessing at the Larmor frequency under the influence of the force $\textbf{F}_M$, which does not require  energy expenditure. Further change in $\textbf{H}$ leads to such a change in $\bm{\omega}_i$, that its final value is always equal to $\bm{o}$. Thus, all bound electrons in the sample with homogeneous $\textbf{H}$ continually and synchronously precess about \textbf{H} at the Larmor frequency. %\footnote{To recall, in the Larmor precession the frequency of orbital revolutions of bound electrons greatly exceeds the precession frequency. In result, neither the orbital radius $R_0$ nor the magnitude of the magnetic moment $|\bm{\mu}_0|$ caused by the orbiting is affected by the field. Accordingly, in a uniform field $\textbf{H}$, the electron orbits precess synchronously retaining the mutual ordering they had at zero field. The condition of the Larmor precession is fulfilled in all magnetics \cite{VK_book}. }. 

Important to note that $r_i$ is proportional but not necessarily equal to the orbital radius $R_0$, it can be either greater or lesser than this. In particular, in many diamagnets $r_i$ is about 1\,$\mathring{A}$, but in bismuth it is close to 4\,$\mathring{A}$ and almost to 10\,$\mathring{A}$ in pyrolytic graphite \cite{VK_JSNM_21}. Therefore, the induced currents can overlap \cite{Ehrenfest}.     

Eq.\,(4) portrays the phenomenon of diamagnetism, an appearance of the field-induced %circular current loops with rms radius $r_i$  and 
microscopic negative magnetic moments, which are unrelated to time and depend neither on temperature, nor on the sign of $e$ and nor on the direction of the orbital magnetic moment at zero field. %$\bm{\mu}$. 
Hence, the  transformation $t\rightarrow -t$ (or $e\rightarrow -e$) makes no difference in magnetization and therefore TRS  holds, in consent  to the Second law. % as applied to diamagnets.  

Now, to superconductors. All properties of these materials (magnetic, electrical and thermal) are determined by Cooper pairs (CPs) formed in statistically significant quantity (typically, on the order of $10^{23}$ per mole\footnote{At $T=0$ all conduction electrons are bound in CPs  \cite{Diego}.}).  In all superconductors, magnetic susceptibility $\chi$ of the \texttt{S} phase is equal to its minimum possible value $-1/4\pi$. %\footnote{For comparison, in normal dia- and paramagnets $|\chi|$ is usually of the order of $10^{-6}$ and $10^{-5}$, respectively.}.  
This means that the magnetic moment of a sample of ellipsoidal shape\footnote{The ellipsoidal shape is a necessary condition for achieving homogeneity of the \textbf{H} field across the sample \cite{Poisson,Maxwell}.} is completely characterized by the field $\textbf{H}$, and, accordingly, its free energy is proportional to $H^2$. Hence, the sample internal energy  is a sum of \textit{kinetic} energies of electrons caused by the field \cite{Tamm,VK_book}, which means that the induced currents \textit{do not interact}.%\footnote{In the case of interacting currents, the sample internal energy would include terms depending on coordinates.}. 
\begin{figure}
	\centering
	\includegraphics[width=0.95\linewidth]{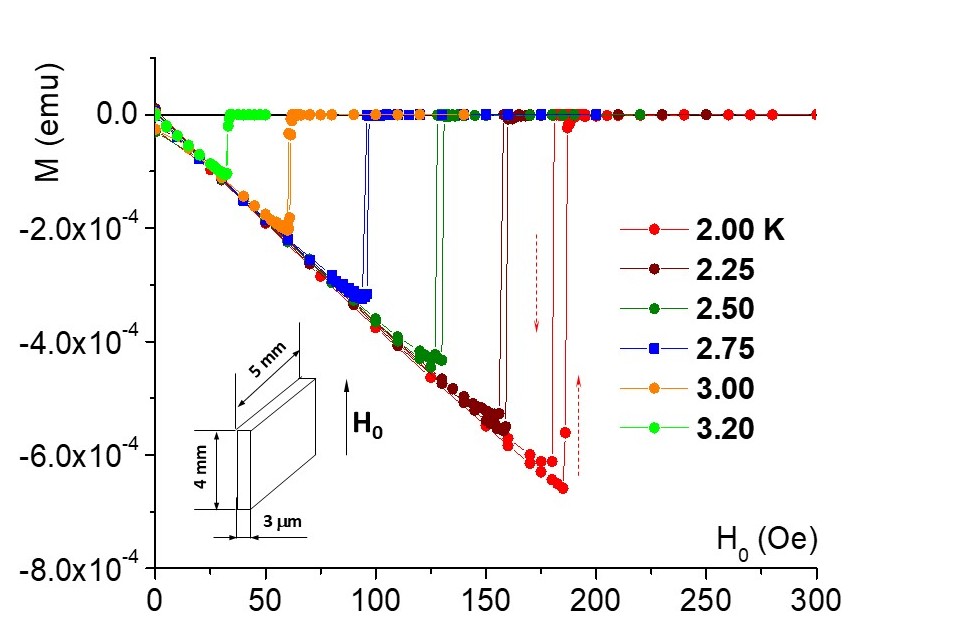} %Here is how to import EPS art
	\caption{Reversible $M$-curves in a type-I superconductor (In film) in the Meissner state, where $M$ is the sample magnetic moment. The insert shows the sample/field configuration; in this geometry $\textbf{H}=\textbf{H}_0$.  The hysteresis at the \texttt{S/N} (\texttt{S} and \texttt{N} stand for superconducting and normal, respectively) transition is the supercooling effect caused by the positive \texttt{S/N} interphase energy (After \cite{VK_JSNM_21}).} 
	\label{fig:epsart}
\end{figure}

CPs represent stable formations of two bound conduction electrons with mutually compensated spins and linear momentums \cite{Cooper}. The first of these conditions ensures stability of CPs (i.e., the thermodynamic advantage of the \texttt{S} state), since, as in normal diamagnets, the free energy of an ensemble of spinless pairs is less than it would be if the pairs possessed spins\footnote{The gain in free energy of superconductors caused by Cooper pairing is well known and easily measurable - this is the condensation energy $E_c=-\int \textbf{M}\cdot d\textbf{H}_0$. }. 

In turn, the second condition implies that paired electrons orbit their centers of mass. Hence, each CP possesses an angular momentum $\bm{\iota}_0$ with a corresponding magnetic moment  $\bm{\mu}_0=\gamma\bm{\iota}_0$, where $\gamma$ is the gyromagnetic ratio of the orbiting electrons, which $g$-factor is equal to unity\footnote{The $g$-factor in a superconductor was for the first time measured by I. Kikoin and Goobar in 1938; they found $g=1\pm 0.03$ \cite{Isaak-1,Isaak-2}.}. Therefore, in the field $\textbf{H}$, pairs precess at the Larmor frequency, and upon relaxation\footnote{In superconductors free of trapped flux, the characteristic time to restore equilibrium  with a change in field or temperature is $\sim (R_0/v_F)\approx 10^{-12}$ s, where $R_0$ is the CP radius and $v_F$ is the Fermi velocity.}, 
each CP acquires the induced moment $\bm{\mu}_{cp}$  equal to\footnote{Derivation of Eq.\,(5) is available in \cite{VK_JSNM_21,VK_book,Diego}} 
\begin{equation}
	\bm{\mu}_{cp}=\frac{\pi r_i^2}{c}\,\frac{(2e)}{2\pi}\bm{\omega}_i=\frac{\pi r_i^2}{c}\,\frac{(2e)}{2\pi}\bm{o}=-\frac{e^2r_i^2}{2mc^2}\textbf{H}. %=-\left(\frac{r_i}{a}\right)^2\frac{\textbf{H}}{4\pi n_{cp}},
\end{equation}

Thus, if the field $\textbf{H}$ is uniform, superconductors free of trapped flux, like regular diamagnets, possess TRS. In terms of magnetization, the main difference between diamagnets and superconductors  lies in the value of $r_i$; in superconductors it is determined by the number density of CPs $n_{cp}$ and is greater by three or more orders of magnitude. Specifically, $r_i^2=m_{cp}c^2/\pi n_{cp}(2e)^2$, where $m_{cp}=2m$ (see, e.g., \cite{Diego}). This leads to $\textbf{I}=n_{cp}\bm{\mu}_{cp}=-\textbf{H}/4\pi$ and, accordingly, to $\textbf{B}=0$. %Respectively, the effects associated with unbound electrons and electrons bound in atoms are negligible. 
An example of fully reversible M-curves of a type-I superconductor in the Meissner state  is shown in Fig.\,1.  

Note that Larmor precession is the \textit{only} kind of electron motion in magnetizing media complying with the Second law, i.e., at which a superconductor (and normal diamagnets) can possess TRS. 
In paramagnets, the mechanism of magnetization %and implementation of TRS 
is different, but it is also based on Larmor precession \cite{VK_book}.

%At the same time, Eq.\,(5) immediately indicates that the entropy of  superconducting samples, both homogeneous and inhomogeneous, in a state of thermodynamic equilibrium is equal to zero. Indeed, the validity of TRS implies that the process of magnetization in these samples is reversible, that is it occurs at constant entropy. On the other hand, the independence of this process on temperature means that that entropy of superconducting samples (in the state of thermodynamic equilibrium) is the same as that at zero temperature. Hence, due to the Third law (Nernst's heat theorem \cite{Nernst,Planck,Schrodinger}), the entropy of superconductors is zero in the entire temperature and field range of the \texttt{S} state\footnote{The magnetic part of entropy of normal diamagnets is also zero. However, due to feeble magnetism and the absent of translational degree of freedom in bound electrons, this fact is insignificant for these materials.}.   

At the same time, Eq.\,(5) immediately evidences that the entropy of superconducting samples in a state of thermodynamic equilibrium is equal to zero. % over the entire range of temperatures and fields of the superconducting state. 
This truly astonishing fact is discussed in the next section.

\vspace{2 mm}

2. Zero entropy. 
\vspace{1mm}

The \textit{key and unique} property of superconductors, dictated by thermodynamics,  is the zero entropy $S$\footnote{The designation of entropy $S$ should not be confused with \texttt{S}, used as an abbreviation for ``superconducting".} of the samples in a state of equilibrium. %Herewith this applies to samples with both homogeneous and inhomogeneous magnetization, that is to those in the Meissner state and in the intermediate and mixed states, respectively. 
Indeed, the induced magnetic moments in all Cooper pairs $\bm{\mu}_{cp}$ in a sample with a uniform $\textbf{H}$ are identical. This fact follows from Eq.\,(5) and is consistent with the cylindrical symmetry of the $\textbf{H}$ %magnetic properties in such 
field. Therefore, the probability $W$ to find this moment in the ensemble of CPs is unity, meaning that its entropy $S=k_B\ln W=k_B\ln 1=0$, where $k_B$ is the Boltzmann constant.

%Indeed, the time-reversal symmetry following  from Eq.\,(5) indicates that the process of magnetization in superconductors occurs at constant entropy $S$. On the other hand, due to identity of the induced magnetic moments  in CPs $\bm{\mu}_{cp}$, following from the 2D symmetry of magnetic properties in the uniform $\textbf{H}$, the probability $W$ of these moments is unity. Hence, the entropy $S=k_B\ln W=0$, where $k_B$ is the Boltzmann constant. % and $N$ is number of CPs in the \texttt{S} phase. 

%On the other hand, the absence of temperature dependence in this process means that $S=S_0$, the entropy at zero temperature. But $S_0=0$ due to the Third law (Nernst's heat theorem \cite{Nernst,Planck,Schrodinger}); hence the above statement follows. 

The fact of zero entropy in superconductors was discovered by Meissner in 1927 \cite{Meissner27}, who revealed the vanishing of the Peltier effect in superconductors. The latter means the vanishing of entropy transport and therefore, taking into account the Third law (Nernst's heat theorem \cite{Nernst,Planck,Schrodinger}), the vanishing of entropy itself. Based on that, Gorter and Casimir adopted zero entropy as the main postulate in their seminal two-fluid model \cite{Gorter_Casimir}. 

Also, the zero entropy in superconductors (and normal diamagnets\footnote{Due to feeble magnetism in normal diamagnets, this fact is insignificant.}) follows from the absence of temperature dependence of magnetization in these materials. Consider, for example, a sample shown in Fig.\,1. %of cylindrical geometry in the parallel field (similar to one shown in Fig.\,1). 
For this case the Maxwell relation reads
\begin{equation}
	\left(\frac{\partial S}{\partial H_0}\right)_T=\left(\frac{\partial M}{\partial T}\right)_{H_0},
\end{equation}
where $M$ is the sample magnetic moment and $T$ is its temperature.

Since $M$ does not depend on $T$, $S$ does not depend on the field. Hence, $S=S_0$, the entropy at zero field.
The latter equals $S_{m0} + S_{lat}$, where the first term is the magnetic component of the entropy at zero field, and the second term is the lattice component, which depends neither on the field, nor on temperature. 
Accordingly, $S_{lat}$ is irrelevant for thermodynamic properties and can be omitted \cite{Schrodinger}.  
On the other hand, by definition, in all diamagnets, including superconductors, $S_{m0}=0$ regardless of temperature. Therefore,  $S=S_0=S_{m0}=0$. 
The same conclusion also follows from  the absence of  the magnetocaloric effect in diamagnets \cite{VK_Thermodynamics}.

The uniqueness of superconductors in the formation of entropy is largely due to %the enormous radius of the induced current loops $r_i$ and 
the mobility of CPs, allowing to achieve a flawless order of the pairs and, consequently,  to minimize the energy of the system. %\footnote{The driving force behind the ordering is the phase gradient discussed below.}. %From the viewpoint of the law of maximum entropy, the fact of zero entropy may seem strange.  However, 

At the same time, zero entropy means realization of what was considered an unrealizable theoretical limit of the Second law: infinite lifespan of the superconducting state simultaneously with zero temperature of the pair ensemble. Therefore, superconductors in the state of equilibrium represent the \textit{ground state} of conducting matter, i.e., the state with the lowest possible energy at a given field. %\tr{Respectively, the internal energy of a  superconductor at zero field represents the zero-point energy.} %As we will see, namely zero entropy leads to one of the most striking properties of superconductors: zero resistance to total current. 

Thus, the zero entropy in superconductors signifies \textit{absolute reversibility} of the properties simultaneously with the \textit{ideal ordering} of induced currents, hereby, this applies to both homogeneous (i.e., Meissner) and inhomogeneous (intermediate and mixed) states. The first point, the reversibility, is consistent with TRS following from Eq.\,(5) and is confirmed by many experiments, e.g., by the data shown in Fig.\,1\footnote{The reversibility of magnetic properties of superconductors was for the first time discovered by Ryabinin and Shubnikov \cite{Shubnikov}.}. Let us see what can be deduced from the second point. % ideal ordering of the induced currents. 

The ideal order implies the maximum symmetry of the circular current loops formed by precessing CPs in a uniform $\textbf{H}$ field. At the same time, as we already know, currents do not interact, which also follows from the fact of zero entropy. One more condition is the homogeneity of induction $\textbf{B}(=0)$. Coming from that, one can figure out the current structure of the  $\texttt{S}$ phase. 

The only way to neutralize the interaction of the current loops is to pile them up one on another in a column. If the column is infinite, i.e., its length is much greater than diameter, and the loops in the column are equally and closely spaced, the column forms an infinite solenoid in which the interaction between the loops is compensated due to the longitudinal symmetry of their location. At the same time, there is no interaction between the columns. The homogeneity of $\textbf{B}$ indicates that the columns are densely packed and therefore make a 2D hexagonal lattice parallel to $\textbf{H}$, which meets the requirement of maximum symmetry. One more consequence of zero entropy is that all induced currents are \textit{in phase}.  

This  qualitative scenario is supported by rigorous reckonings based on Bohr-Sommerfeld quantization condition applied to Cooper pairs. The corresponding model is called the micro-whirls model (MWM), where the micro-whirls represent infinite solenoids tightly ``wound" about the \textbf{H} lines\footnote{The longitudinal spacing between the current loops is a universal constant equal to $2e^2/mc^2\approx 5.6\, fm$, about a tripled diameter of the proton. }. %; the model is free of postulates. 
MWM consistently describes \textit{all} known properties of superconductors and predicts some novel  effects\footnote{In particular, the possibility of two electron-spin resonances with $g$-factors equal to 2 and 1, due to unbound or free electrons and electrons bound in CPs, respectively.}.  The model is discussed in detail in \cite{VK_book} and in a shortened form in \cite{Encyclopedia}. Basics of MWM along with the experimental data that prompted its development are available in \cite{VK_JSNM_21}. The idea of micro-whirls was first proposed by Hall \cite{Hall}. To illustrate, a cross section of the micro-whirls structure is schematically shown in Fig.\,2.

\begin{figure}
	\centering
	\includegraphics[width=0.95\linewidth]{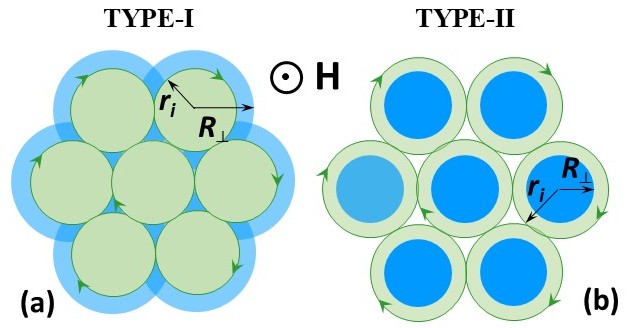} %Here is how to import EPS art
	\caption{The micro-whirls structure of the Meissner state in type-I (a) and type-II (b) superconductors. The radius $R_\perp=\sqrt{(2/3)}R_0$, where $R_0$ is the radius of the orbital motion of electrons bound in CPs \cite{Diego}.  In (a) $r_i<R_\perp$, which leads to positive interphase energy in samples with  inhomogeneous magnetization; in (b) $r_i>R_\perp$ and the interphase energy in such samples is negative.  The cylinders of  radius $R_\perp$ (shown in blue) are formed by Cooper pairs of all orientations constituting a given micro-whirl.  
	The cylinders of radius $r_i$ (shown in green) are the micro-whirls formed by the induced currents shown by solid green circles with arrows; all of them are in phase. Radii $r_i$ and $R_\perp$ represent root-mean-square values, so the currents partially overlap not leaving voids. The field $\textbf{H}$ is directed toward the reader. } 
	\label{fig:epsart}
\end{figure}   

On the other hand, the zero entropy means that temperature of the ensemble of ordered CPs ($T_{cp}$), i.e.,  of the micro-whirls, is zero regardless on temperature of the lattice \cite{Nernst}. In turn, zero $T_{cp}$ means that  micro-whirls do not interact with the environment. Hence, in the subcritical total current the micro-whirls move relative to the lattice with a drift velocity ${v}_{dr}(\ll\omega_i r_i) $  experiencing \textit{no resistance}, i.e., maintaining the order of the micro-whirls' lattice\footnote{In supercritical current, CPs are disordered, leading to a resistive state, which resistance is lower than that of the \texttt{N} state \cite{Shubnikov-Alexeyev,Scott,Mukherjee}.}. Note that the latter also follows from the Second law (see below).  

%After all, the absence of entropy means that properties of superconductors in equilibrium states are conditioned exclusively by intrinsic properties of their material. 

\vspace{2mm}
3. Two kinds of superconducting currents. 
\vspace{1mm}

In normal metals, there are two kinds of currents: dissipation-free molecular  currents of bound electrons and dissipative total currents of unbound electrons flowing along macroscopic paths \cite{Tamm}. Electrons forming these currents behave so differently that it is pertinently to speak about different electrons.  

This is not the case in superconductors: all currents relevant to superconductivity are due to conduction electrons bound in CPs. However, currents in superconductors are also of two kinds: circular currents due to precessing CPs, which form micro-whirls, and the total currents due to translational motion of the whirls. From a viewpoint of thermodynamics, the principal difference between these currents is that the former determine equilibrium properties, while the latter - nonequilibrium ones. 

To recognize the existence of such currents, it is advisable to consider the quantum-mechanical expression for the current density $\textbf{j}$ in a magnetic field (see, e.g., \cite{Landau_QM}).   
For spinless CPs with mass $2m$, charge $2e$ and the number density $n_{cp}$, this expression reads 
\begin{equation}
	\textbf{j}= \frac{ (2e)}{(2m)}\left[\frac{i\hbar}{2}(\Psi\nabla\Psi^*-\Psi^*\nabla\Psi)-\frac{2e}{c}\textbf{A}\Psi\Psi^*\right],
\end{equation}
where $\Psi=n_{cp}^{1/2}e^{i\varphi}$ is the wave function of CPs with the phase $\varphi$, where $\hbar$ is the reduced Planck constant, $i$ is the imaginary unit, and $\textbf{A}$ is the vector potential experienced by  electrons belonging to the sample, i.e.,  the vector potential  of the field $\textbf{H}$. 

After substituting $\Psi$, Eq.\,(7) takes the form
\begin{equation}
	\textbf{j}=\frac{\hbar n_{cp} (2e) }{(2m)}\left[\nabla\varphi-\frac{2e}{c\hbar}\textbf{A}\right].%\qquad (A5.2)
\end{equation}  

%Comparing $\textbf{j}$ with the general formula that following from the continuity equation
Comparing Eq.\,(8) with $\textbf{j}=(2e)n_{cp}\textbf{v}_{cp}$ (continuity condition), 
one finds a linear velocity of CPs, which is
\begin{equation}
	\textbf{v}_{cp}=\frac{\hbar}{2m}\nabla\varphi-\frac{e}{mc}\textbf{A}=\frac{\hbar}{2m}\nabla\varphi-\frac{e}{2mc}\textbf{H}\times \textbf{r}_i.
\end{equation}

\begin{figure}
	\centering
	\includegraphics[width=0.88\linewidth]{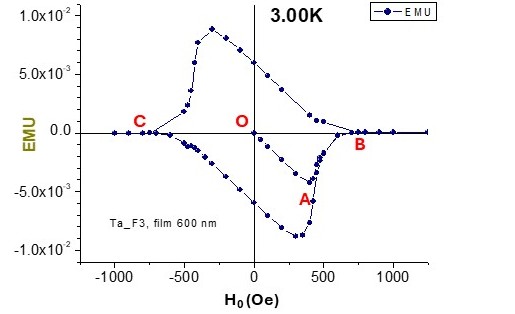} %Here is how to import EPS art
	\caption{Magnetic moment of a 600-nm-thick Ta film %of ($4 \times 6\times 6\cdotp10^{-4}$ mm$^3$) 
		in a parallel field $\textbf{H}_0$. The film was cooled at zero field (p.\,O). In OAB the film is in thermodynamic equilibrium: in the Meissner state over OA and in the mixed state over AB. In all other sections  below the critical field $H_{c2}$ (pp.\,B and C) the film contains total currents and is out of equilibrium. } 
	\label{fig:epsart}
\end{figure}

Here the second term is the linear velocity of the field-induced \textit{circular motion} of the paired electrons $v_i(=\omega_ir_i= eHr_i/2mc)$ caused by the Larmor precession of CPs and leading to the perfect diamagnetism of the ensemble of ordered pairs, and the first term is the velocity of \textit{translational motion} of CPs driven by the phase gradient. Accordingly, the second term portrays microscopic circular currents, while the first one represents the total current of the pairs\footnote{The same follows from MWM, where electron motion is a superposition of the circular motion in the whirls and, in the presence of the total current, the translational motion with the whirls.}.   A well-known example of the total current caused by the phase gradient is the Josephson effect \cite{Josephson-62}. 
Note that if $\nabla\varphi=0$, then the total current is absent, as required by the Second law. This confirms that  CPs precess synchronously.

There are also two more sorts of the total current: transport current in a circuit with a superconducting section and a field-induced current in multiply-connected superconductors. The former is due to external e.m.f. and the latter is caused by induced e.m.f. conditioned by the flux trapped in the sample opening(s).

These kinds of the total current were first revealed by Meissner and Ochsenfeld in their arrangement with an external e.m.f. \cite{Meissner} and by Shoenberg in an experiment with a ring %\footnote{The aim of Meissner and Ochsenfeld in the experiment with the transport current was to determined either the current flows on the surface or in the bulk; they found that it flows in the bulk. On the contrary, Shoenberg found that current in the ring flows on the surface. In both cases it was the total current which flows without distraction of the whirls' order \cite{VK_book}.}
\cite{Shoenberg}. Note that simply-connected samples with the flux trapped in pinning centers represent %are identical to 
multiply-connected bodies since the trapped flux is always surrounded by the total current (the Faraday law). Magnetic moments due to these currents are determined by the currents geometry and by the sign of the applied field \textit{change}. Specifically, the moments are negative when $\textbf{H}_0$ increases and positive when it decreases, whereas the moments due to intrinsic currents are always negative, as is seen from Eq.\,(5). An example of the M data obtained on a simply-connected sample with and without total currents is shown in Fig.\,3. For comparison, the field dependence of the magnetic moment of a multiply-connected sample (a ring) is shown in Fig.\,4.        

\begin{figure}
	\centering
	\includegraphics[width=0.7\linewidth]{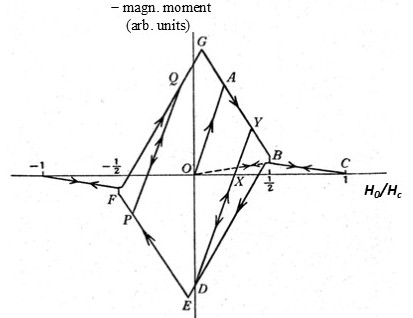} %Here is how to import EPS art
	\caption{Magnetic moment of a superconducting Pb ring measured vs $\textbf{H}_0$ at constant temperature; $\textbf{H}_0$ is applied transverse to the ring plane. The ring was cooled at zero field (p.\,O); a dashed section OB represents the equilibrium magnetic moment (the moment that the ring would have in the absence of the total current), which is completely overshadowed by the moment caused by total current; there is no total current in the section BC, where the ring is in the intermediate state, where it consists of alternating $\texttt{S}$ and $\texttt{N}$ layers perpendicular to the total current. After Shoenberg \cite{Shoenberg}. }
\end{figure}

Any total current represents a transport phenomenon. Hence, samples with total currents are out of thermodynamic equilibrium and, consequently, do not possess TRS. Therefore, according to the Second law, the properties of such samples must evolve toward the state of equilibrium. This may sound strange since usually when a superconductor is kept at fixed external conditions its properties do not change. However, in reality the total current does decay, albeit at a very slow rate; e.g., in a superconducting magnet made of Nb/Zn wire, the field of 2 kG decays with a rate $\sim 10^{-6}$ G/h, meaning that the total current ceases in 100 thousand years\footnote{Note a gargantuan difference in relaxation rates in superconductors without (see footnote ($^{17}$) above) and with total current.} \cite{File}.

At the same time, the decay of the total current \textit{cannot} be accompanied by the disordering of CPs, since otherwise the entropy of CP's ensemble would become nonzero, and, consequently, the system would not be able to return to the ordered state if the total current vanishes on some reason, e.g., due to cutting the ring in Fig.\,4  or due to the passage of sufficient time. 

Therefore, the total current decays due to radiation, since it flows through a closed path. The apparent slowness of this process is the origin of the phenomenon called persistent current. It is this factor that enables the enormous  advantages of superconductivity in applications. Accordingly, if the total current  is switched off or vanishes itself,  the sample becomes a disconnected ZFC superconducting wire in a magnetic field (which magnetic moment is equal to that at a point in the section OB in Fig.\,4), or a ring with no trapped flux, respectively. Consequently, the entropy of the sample free of the total current remains zero as required by the Second law.  
%Respectively, if the total current  is switched off,  the sample magnetic moment becomes equal to that at a point in the section OB (Fig.\,4),  since the ring becomes a disconnected ZFC superconducting wire in a magnetic field, or into a ring with no trapped flux. 

\vspace{2mm}
4. Necessary conditions for achieving thermodynamic equilibrium in superconductors.  

\vspace{1mm}

%Normally, samples used in physics research are in a state of thermodynamic equilibrium (unless the aim is to study transport properties).  
Samples used in studies of intrinsic properties of materials should be in a state of thermodynamic equilibrium. 
But achieving equilibrium in superconductors can seem as an extremely difficult task, especially for multicomponent  compounds, such as, e.g., high-$T_c$ materials.
Actually, however, there are just two quite feasible conditions, the fulfillment of which is necessary to attain  equilibrium in \textit{any} superconductor. These conditions are as follows.

(1) \textit{Homogeneity of the field strength} $\textbf{H}$, which means that the sample shape should be close to ellipsoidal. Otherwise, the $\textbf{H}$ field and, accordingly, the Larmor frequencies will be different throughout the sample, which will lead to the appearance of phase gradients  and, respectively, to uncontrollable  total currents. 

(2) \textit{Absence of the trapped flux}. % lines. 
Excluding rare fully reversible cases, this condition means that the sample should be cooled at zero field. Herewith, to prevent occasional artifacts,  it is safer to vary the applied field in one direction. 
At this condition the trapped flux is negligible and, consequently, the sample properties are determined by the intrinsic properties of its material (see, e.g., \cite{Diego}).  

Note that there is no requirement for purity. It is desirable, but not a priority. If the above conditions are not met, the purity is irrelevant at all. As an example, one can cite measurements of the Barnett effect in ZFC disk-shaped superconducting samples: Pb (poly-crystal), YBa$_2$Cu$_3$O$_{7-\delta}$ and BPb$_{0.8}$Bi$_{0.2}$O$_3$ (both ceramics); in spite of the abundance of defects in the ceramic samples, the results obtained after correction for the volume of the \texttt{S} phase  were the same    \cite{Verheijen-90}.
\vspace{2mm}

Summary and outlook. 
\vspace{1mm}

The features of the Second law %of thermodynamics 
as applied to superconductors were considered, with particular attention to the concepts of the equilibrium states and the reversibility of thermodynamic processes in these materials. The main of these features include (1) the zero entropy of the ensemble of Cooper pairs in samples that are in thermodynamic equilibrium; and (2) the existence of two kinds of superconducting currents in samples that are out of equilibrium.

%The features include (1) the primary role of the field strength $\textbf{H}$ in the formation of the properties controlling superconductivity; (2) the zero entropy of the ensemble of Cooper pairs in samples that are in thermodynamic equilibrium; and (3) the existence of two kinds of superconducting currents in samples that are out of equilibrium. 

Necessary conditions for attaining  thermodynamic equilibrium in superconductors were framed out, namely: an ellipsoidal shape of samples % used in studies of superconductivity; 
and the zero-field cooling, desirably with compensated terrestrial field. Sample purity is important, but is not the primary objective in preparation of the experiment; if the above conditions are not fulfilled, purity is of no significance.

It was shown that the Second law justifies and explains the phenomenon of superconductivity. In particular, the Second law addresses the questions \textit{why} it occurs [due to thermodynamic profitability] and \textit{how} does it work [via Cooper pairs organized into the micro-whirls' structure, due to thermodynamic advantage of such a structure]. Hence,  unlike theories considering a genesis of electron pairing in specific lattice environments, the thermodynamic analysis sheds light on what can be called as the purpose (pursued by nature) in the formation of the dissipation-free state of charge carriers and the mechanism that allows its achievement.
The most important property of superconductors, determining %all 
other properties of these materials, is zero entropy of thermodynamically equilibrium samples. In turn, the zero entropy means infinite lifespan of the superconducting state. Therefore, superconductors in the state of equilibrium represent the ground state of conducting matter.

Evidently, however, that provided analysis is not the end of the story. In particular, the Second law suggests that any metal should be superconducting below the critical temperature of a given material.  %There is hardly a reason to doubt this prediction. 
However, despite the prevalence of superconductors with diverse chemical and structural compositions, not all of them, including some of ``the best" ones (alkali metals)  have been observed in this state
\cite{Schilling,Schaeffer}. One can speculate that elucidating the mechanism(s) \textit{preventing} these metals from becoming superconducting may provide the key to controlling superconductivity. 

Returning to the problem of reversibility, the reversible processes were introduced and always considered as a theoretical idealization. Citing Planck, ``They are, however, of considerable importance for theoretical demonstration and for application to states of equilibrium" \cite{Planck}. Yet, in superconductors such processes are \textit{real and readily observable} (providing the samples are in the state of equilibrium), and their properties are fully consistent with requirements of the Second law. Thus, the superconductivity phenomenon can be viewed as one more  triumph for classical thermodynamics.
\vspace{2 mm}

\textbf{Acknowledge.} I am deeply grateful to professor Oscar Bernal for criticism and valuable comments on the manuscript.

\end{document}